\documentstyle[prl,aps,multicol,epsf]{revtex}
\begin{document}
\sloppy
\title
{\bf {The dynamo effect - a dynamic renormalisation group approach}}
\author{Abhik Basu$^1$ and Jayanta K Bhattacharjee$^2$, \\
$^1$Department of Physics, Indian Institute of Science, \\Bangalore 560012 
INDIA,\\
and $^2$Department of Theoretical Physics,\\ Indian Association of Cultivation
of Science, \\ Calcutta 700032, INDIA.}
\maketitle


\begin{abstract}The Dynamo effect is used to describe the generation of 
magnetic fields in astrophysical objects. However, no rigorous derivation
of the dynamo equation is available. We justify the form of the equation
using an Operator Product Expansion (OPE) of the relevant fields. We also
calculate the coefficients of the OPE series using a dynamic renormalisation
group approach and discuss the time evolution of the initial conditions
on the initial seed magnetic field.
\end{abstract}


The existence of magnetic fields in the astrophysical objects has been an 
active area of research for quite sometime. This effect is known as
the dynamo effect. Many mechanisms have been suggested to explain this.  
The dynamo effect could be caused by a background turbulent fluid (see, 
for example [1]) or by a non-turbulent backgorund fluid motion 
(an interesting example of this type has been discussed in [2]). In this letter
we, however, confine ourselves to the study of the first kind. 

  In turbulent dynamo, to begin with one assumes that initial seed field 
is very weak i.e. most of the energy is contained in the kinetic part of the
magnetised fluid. This immediately tells us that even though the velocity
field is influencing the time evolution of the magnetic field, the back
reaction of the magnetic field on the velocity field is negligible as the
ratio of the 
the Lorentz force to the inertial force 
is $\sim B^2/u^2$ (this is an order of magnitude
estimate of the two nonlinear term in the Navier-Stokes equation) 
where $\bf B$ and $\bf u$ are the
magnetic and velocity fields. Basically, the velocity field being
unaffected by the magnetic field, will time
evolve according to the Navier-Stokes (NS) equation and magnetic field will
time evolve according to the Induction equation, being 
influnced by the velocity
field. However, this assumption of weak magnetic field will hold only
during the initial transient. 

Conventionally, while deriving the dynamo equation one takes a two
scale approach [1]: It is assumed that there are two scales of variations
-(i) A global scale of variation $L$ of 'mean' quantities and (ii)a scale of
small wavelength fluctuations $l_o$. Mean quantities are defined to be
quantities which are averaged over an intermediate scale $a$:$l_o<<a<<L$.
In other words, fields have a small wavenumber $\sim L^{-1}$ and a large
wavenumber $\sim l_o ^{-1}$ components and the mean quantites can be
calculated by averaging over the large wavenumbers. We split the velocity
and the nagnetic fields into mean and fluctuating parts:
\begin{eqnarray}
{\bf U}({\bf x},t)={\bf U_o}({\bf x},t)+{\bf u}({\bf x},t),<{\bf u}>=0 \\
{\bf B}({\bf x},t)={\bf B_o}({\bf x},t)+{\bf b}({\bf x},t),<{\bf b}>=0
\end{eqnarray}
$< >$ indicates an averaging over $a$, the intermediate scale.

The time evolution of the magnetic fields is governed by the Induction
equation [1,3]:
\begin{equation}
{\partial {\bf B}\over\partial t}={\bf \nabla}\times ({\bf u}\times
{\bf B}) + \mu \nabla^2 {\bf B}
\end{equation}
where $\mu$ is the magnetic viscosity. 
With the above definitons of the mean and fluctuating quantities, one can
easily write down the equations for $\bf B_o$ and $\bf b$. The one for
$\bf B_o$ is given by
\begin{equation}
{\partial {\bf B_o}\over\partial t}={\bf \nabla}\times ({\bf U_o}\times 
{\bf B_o}) + {\bf \nabla }\times {\bf E} +\nu \nabla ^2 {\bf B_o}
\end{equation}
where ${\bf E}=<{\bf u}\times {\bf b}>$. It is easy to see that, even though
$<{\bf u}>$ and $<\bf {b}>$ are zero individually, their product may yield
a non-zero value because of a possible long range correlation between the
two. Then a gradient expansion is performed over 
this {\it electromotive force} $\bf E$.
This has been justified by the argument that $\bf E$ and $\bf B_o$ are
linearly related:
\begin{equation}
E_i=\alpha_{ij}{B_o}_{j}+\beta_{ijk}{\partial {B_o}_j\over\partial x_k}
+\gamma_{ijkl}{\partial ^2 {B_o}_j\over \partial x_k \partial x_l}+...
\end{equation}
In other words, averaging $\bf u\times b$ over the intermediate scale
gives rise to a gradient expansion of the long wavelength part of
the magnetic field. The first term is 
responsible for $'\alpha$-effect' and the second term 
for $'\beta$-effect'. It can be easily shown that when the background 
fluid motion is homogenous and isotropic, $\alpha$ is a pseudo scalar 
and $\beta$ is a true scalar. The $\beta$ term contributes to the effective
viscosity (`turbulent diffusion'). Hence, including these $\alpha$ and
$\beta$ effects, the equation for the large scale component of the
magnetic fields becomes
\begin{equation}
{\partial {\bf B_o}\over\partial t}={\bf \nabla}\times ({\bf U_o}\times
{\bf B_o})+\alpha{\bf \nabla}\times\ {\bf B_o}+\eta \nabla^2 {\bf B_o}
\end{equation}
where $\eta=\nu+\beta$, the effective viscosity. This is the dynamo equation.

Even though this equation has been very successful, the justification for
the existence of a gradient expansion (which is a crucial part in the
derivation) is not founded upon a strong basis.
Here we attempt to give a rigorous basis to it: From a field theoretic
point of view $<{\bf u(r_1)}\times {\bf b(r_2)>}$ (here $\bf u(r_1)$ and
$\bf b(r_2)$ are the 'fluctuating' part of the total velocity and
the magnetic field as defined in equations (1) and (2)) 
diverges as $r_1\rightarrow
r_2$. Hence, one can write down an Operator Product Expansion (OPE) [4] for
the product of the fields $Lt_{r_1 \rightarrow r_2} u(r_1) \times b(r_2)$
and consequently calculate $Lt_{r_1\rightarrow r_2} <{\bf u}\times {\bf b}>$.
Since the LHS is linear in $\bf b$, the RHS will have only odd powers
of $\bf B_o$ (so that $\bf B\rightarrow -B$ property 
is retained). For the same
reason, there cannot be any term having no $\bf B_o$. Also, no term 
propertional to $\bf b$ can appear as $\bf b.u\times b=0$ (the LHS). 
However,  no such condition is there for the velocity field as if $\bf u$
is a solution of the NS equation then $-\bf u$ is not.
We write down the OPE as
\begin{equation}
({\bf u\times b})_i \equiv E_i = \alpha_{ij}{B_o}_j+\beta_{ijk}{\partial 
{B_o}_j\over\partial x_k}+...+[]
\end{equation}
where [] indicates various 
composite operators permitted by the symmetry
of the LHS. Since, expectation value of any composite operator is 
defined to be zero we have
\begin{equation}
<{\bf u\times \bf b}>=\alpha_{ij} {B_o}_j +\beta_{ijk} {\partial {B_o}_j
\over\partial x_k} +....
\end{equation}
which is in agreement with the form of the equation.

In $k$-space, $Lt_{r_1\rightarrow r_2}<\bf u\times\bf b>=\int_q {\bf u}(q)
\times {\bf b}(k-q)$. Here, we take $k$ to be small i.e. long wavelength
limit and $\Lambda >q>\Lambda e^{-r}$ with $r$ positive, i.e. $q$
belongs to some short wavelength band. The RHS becomes, $\int_q\alpha_{ij}(q) {B_o}_j (k-q)
+\int_q\beta_{ijk}(q) k_j {B_o}_k (k-q)+...$ (in principle, $\alpha,\beta$
etc. could be functions of $k$). For calculating the OPE coefficients, we
calculate the LHS by momentum shell integration, do a loop expansion
and equate various powers of $k$ on both sides. We work in the incompressible
limit.

The Navier-Stokes equation is given by
\begin{eqnarray}
{\partial {\bf u}\over\partial t}+(\bf u .\nabla){\bf u}=\nu\nabla^2 {\bf u}
-{\nabla p\over\rho} +\bf f \\
\nabla .\bf u=0 
\end{eqnarray}
The last one is the incompressibility assumption. 
Here, $\nu$ is the fluid viscosity, $p$ the pressure and $\rho$ the
density. This in $k$-space reduces to 
\begin{equation}
{\partial u_{\alpha} (k,t)\over\partial t} +{i\over\ 2}P_{\alpha\beta\gamma} 
(k) \int_q 
u_{\beta\gamma} (q) u_k (k-q) =-\nu k^2 u_{\alpha} +f_{\alpha}
\end{equation}
where $P_{\alpha\beta\gamma}=P_{\alpha\beta}k_{\gamma} 
+ P_{\alpha\gamma}k_{\beta}$ and $P_{\alpha\beta}$ is the projection
operator. Since we are looking at turbulent dynamo, 
we take $<f_{\alpha}(k,t)f_{\beta}(k^{\prime},
t^{\prime})>=D_1 (k) P_{\alpha\beta} 
\delta (k+k^{\prime})\delta(t-t^{\prime})
+D_2 (k)\epsilon_{\alpha\beta\gamma} k_{\gamma}
\delta(k+k^{\prime})\delta(t-t^{\prime})$.
$D_1$ and $D_2$ are even in $k$.
An explicit factor of $i$ infront of $D_2$ term indicates that it is an
odd parity breaking term and according to our previous
analysis $\alpha$-effect will be absent if $D_2=0$.  

The induction equation is given by
\begin{equation}
{\partial {\bf B} \over\partial t} =\bf \nabla \times (\bf u\times b) +
\mu \nabla ^2 {\bf b}
\end{equation}
This, in $k$-space becomes 
\begin{equation}
{\partial b_{\alpha}(k,t)\over\partial t}=i\epsilon_{\alpha\beta\gamma}
k_{\beta} \epsilon_{\gamma\mu\lambda}
\int_q u_{\mu} (q) b_{\lambda} (k-q) -\mu k^2 b_{\alpha}
\end{equation}

At the tree level (fig.1), the diagram which will contribute is the following:
\vspace{5cm}
We have, $<({\bf u\times \bf B})_{\alpha}>=<i\int_{q,\Omega}\epsilon_{\alpha
\beta\gamma}\epsilon_{\gamma\delta\lambda}\epsilon_{\lambda\eta\tau})
(k-q)_{\delta}
[D_1 (q)P_{\beta\eta}(q)+D_2 (q)\epsilon_{\beta\eta\rho} q_{\rho}
] 
G_o ^b ( k-q,\omega-\Omega) C_o ^u( q,\omega) (k-q)_{\delta}
B_{\tau} (k) $, where $G_o ^b (q,\Omega)={1\over\ i\Omega +\mu q^2}$
and $C_o ^u (q,\Omega) ={1\over\ \Omega ^2 +\nu^2 q^4}$. We extract the
$\alpha$-term and the $\beta$ term from the above expression in the
following way:

{\it I. The $\alpha$ term}: In the above integral, there is a term which is
independent of $k$. This gives rise to the $\alpha$ term
\begin{equation}
\alpha B_{\alpha}(k)= 
\int d^3 qd\Omega \epsilon_{\alpha\beta\gamma}\epsilon_{\gamma
\delta\lambda}\epsilon_{\lambda\eta\tau}\epsilon_{\beta\eta\rho}
{(k-q)_{\delta} q_{\rho} D_2 (q) \over\ \Omega^2 +\nu^2 q^4}
{1\over\ i(\omega-\Omega)+\mu (k-q)^2} B_{\tau}(k)
\end{equation}
Now, $\epsilon_{\alpha\beta\gamma}\epsilon_{\gamma\delta\lambda}
\epsilon_{\lambda\eta\tau}\epsilon_{\beta\eta\rho}
=\delta_{\alpha\delta}\delta_{\tau\rho}+\delta_{\alpha\rho}\delta_{\tau
\delta}$. Hence, one obtains
\begin{equation}
\alpha={4\pi^2\over\ 3}{1\over\nu(\nu+\mu)}\int_q D_2 (q)
\end{equation}
Since, the integral is infra-red divergent, we evaluate it by shell 
integration.

{\it II. The $\beta$ term}: The term proportional to $k$ gives rise to the
$\beta$ term
\begin{equation}
i\beta\epsilon_{\lambda\delta\tau}=
i\int d^3 qd\Omega D_1 (q) \epsilon_{\alpha\beta\gamma}
\epsilon_{\gamma\delta\lambda}\epsilon_{\lambda\eta\tau}P_{\beta\eta}(q)
(k-q)_{\delta}
\end{equation}
Now, $\epsilon_{\alpha\beta\gamma}\epsilon_{\gamma\delta\lambda}
=\delta_{\alpha\delta}\delta_{\beta\lambda}-\delta_{\alpha\lambda}
\delta_{\beta\lambda}$. Hence,
\begin{equation}
\beta={4\pi^2\over\ 3}{1\over\nu(\nu+\mu)}\int_q {D_1 (q)\over\ q^2}
\end{equation}
\noindent Notice that $\alpha$ is pseudo-scalar and $\beta$ is a true scalar.
We see that even for thermal noise i.e $D_1$ and $D_2$ being constants
give nonvanishing $\alpha$ and $\beta$. However, if we assume that
the background fluid is fully developed turbulent, then $D_1(k)\sim k^{-3}$
and $D_2 (k) \sim k^{-4}$ (this particular choice gives $E(k) \sim k^{-5/3}$
[5]).

Having calculated $\alpha$ and $\beta$ from the tree level diagram, we now
analyse the higher-loop diagrams. There are several distinct classes of them:
(i) This type can be generated by decorating the velocity correlator or
the response function. They generate corrections to the coefficients already
obtained without decorations (fig 2a). (ii) This type of diagrams are 
generated by decorating the magnetic field response function.  They give
rise to non-zero values for the higher order coefficients of the OPE
series as well as corrections to the coefficients already obtained 
without decorations (fig 2b). 
These two classes of diagrams can be obtained by
considering a 'self-consistent' tree level diagram.

(iii) These type of diagrams appear by inserting two ub vertex (the advective
vertex of the Induction equation) in the tree level diagram (fig 3a) or
by inserting one uu and one ub vertex (fig 3b and 3c).
A 1-loop diagram as shown in fig 3a is given as
$\int_{q,p}G_o ^b (k-q) C_o ^u(q) A_1(p)$ 
where 
\begin{equation}
A_1(p) =G_o ^b (k-q-p)G_o ^b(k-p)C_o ^u(p)
\end{equation}
Similarly, fig 3b and fig 3c are given by
$\int_{q,p} G_o^b(k-q)C_o ^u(q) A_2(p)$ and 
$\int_{q,p} G_o^b(k-q)C_o^u(q) A_3(q)$
where
\begin{eqnarray}
A_2(p)=G_o^b(k-p)G_o^u(p-q)C_o^u(p) \\
A_3(p)=G_o^b(k-p)G_o^u(-p)C_o^u(p-q)
\end{eqnarray}
We see that the sum of $A_1,A_2$ and $A_3$ are nothing but the 1-loop 
correction of the advective vertex of the Induction equation. Sum of
$A_1,A_2$ and $A_3$ are however not 
zero as we are calculating the vertex correction
at finite external momenta. These diagrams give rise to 1-loop corrections
to $\alpha$ and $\beta$ (which are already present in the tree level) as well 
as the lowest order $\gamma$.
It is easy to see that an $n$-loop diagram will give rise to terms upto the
$n$-th term in the OPE series starting from $\alpha$. 

A pertinent question at this stage is:How does initial correlations among
the seed magnetic field (a dynamo is essentially an initial value problem)
grow?
Also, if the initial magnetic field correlations are parity symmetric, will
it generate a parity breaking correlation after a finite time (possibly
due to the
parity breaking correlation of the forcing in the Navier-Stokes equation)?
Consider a 1-loop diagram which would renormalise the initial magnetic field
correlations: It is clear that inside such a 1-loop integral, one noise 
correlation arises from the NS equation and the second one is from the
initial magnetic field correlation. Hence, if the N-S noise has a 
nonvanishing parity breaking term, it will give a non-zero contributuion
to the 1-loop integral only if the initial seed magnetic field correlation
also has a parity breaking term, as one requires two or even number 
`epsilon' terms for the integral not to vanish. Consequently, we can say
that unless the initial correlation has a parity breaking term, it will
not be generated in the future time. It is also clear that even if the
intial magnetic field correlation is non-singular at $t=0$, it will pick up
a singular contribution at a later time (since the noise in the NS equation
has a singular correlation).

It is clear that due to the dynamo action, the initial magnetic field will
grow. Now, once $\int B^2 /\int v^2$ becomes $\sim 1$, it is nolonger
justified to neglect the Lorentz force term, which we have neglegted so far.
Adding this term to the N-S equation we have the following set of equations:
\begin{eqnarray}
{\partial {\bf u}\over\partial t} +{\bf (u.\nabla)u}=\nu\nabla^2 {\bf u}+
+ {1\over\ 4\pi\rho}(\nabla\times {\bf B)\times B} +{\nabla p\over\rho} +
{\bf f} \\
\nabla .\bf u=0 \\
{\partial \bf B\over\partial t}=\alpha \nabla \times {\bf B} +
\nabla \times ({\bf u\times B})+\eta \nabla ^2 {\bf B} \\
\nabla .\bf B=0
\end{eqnarray}
We choose
\begin{equation}
<f_i(k,t)f_j(k^{\prime},t^{\prime}>=[D_1 P_{ij}k^{-3}+iD_2 \epsilon_{ijp} k_p]
\delta(k+k^{\prime})\delta(t-t^{\prime}) 
\end{equation}
\begin{equation}
<B_i(k,t=0)B_j(k^{\prime},t=0)>=DP_{ij}k^{-3}\delta(k+k^{\prime})
\end{equation}
It is clear the dynamo equation has two linear terms - 
one proportional to $k$ and the other
proportional to $k^2$. Conventionally, one may like to keep all the linear
term in the bare response function. However, that would give rise to a 
singularity at a finite real $k$. To avoid that problem, we keep only
the viscosity term in the bare response function and treat the $\alpha$
term perturbatively. Following Forster {\it et al} [6] we obtain the
following recursion relations for the parameters:
\begin{eqnarray}
{d\nu\over\ dr}=\nu(r)[z-2+A_1 {D_1\over\nu^3 \Lambda^4} 
+A_2 {D_2\over\nu\eta^2 \Lambda^4}] \\
{d\eta\over\ dr}=\eta(r)[z-2+A_3 {D_1\over\nu(\nu+\eta)\Lambda^4}
+A_4 {D_2\over\eta(\eta+\nu)\Lambda^4} +A_5 {\alpha^2\over\eta\Lambda^2}] \\
{d\alpha\over\ dr}=[z-1+A_6{1\over\alpha\nu(\nu+\eta)\Lambda^{1/3}}]
\end{eqnarray}
One can see easily that $\alpha \sim k^{-1/3}$ and $\eta \sim k^{-4/3}$.
$A_1,...,A_5$ are numerical constants which determine the non-universal
amplitudes.
It is clear that all the exponents dynamic (=2/3) and roughness exponents
are same as usual MHD turbulence. Only the nonuniversal amplitudes are
different. Here $\eta$ is the effective viscosity (or the 'turbulent
diffusion' as opposed to the bare molecular viscosity. This is the origin
of the enhanced diffusion process in many astrophysical situations e.g
accretion disk [7].

So, in conclusion, we have shown how the dynamo equation can be formally 
constructed using OPE. We have shown how to calculate the coefficients
of the OPE series pertubatively in a loop expansion. In particular we
have evaluated $\alpha$ and $\beta$ in a momentun shell ellimination
method. We have also discussed renormalisation
of the dynamo equation.


\newpage
\noindent Figure Captions:
Fig 1: Tree level diagrams for $<\bf u(q)\times b(k-q)>$. A solid line
indicates a bare magnetic field response function, a broken line indicates
a bare velocity response funtion, a 'o' joined by two broken lines indicates
a bare velocity correlation function, a wavy line indicates a magnetic field,
a solid triangle indicates a $ub$ vertex.

Fig 2a and 2b: 1-loop corrections to the previous tree level diagram which 
arise due to dressing of the bare velocity response/correlation functions
and bare magnetic field response function, a 'X' indicates a $uu$ vertex.
Other symbols have same meaning.

Fig 3a, 3b and 3c: 1-loop correction to tree level diagram arising out of
1-loop corrections to the $ub$ vertex. Symbols have same meaning.

\begin{figure}[htb]
\epsfysize=10cm
\centerline{\epsfbox{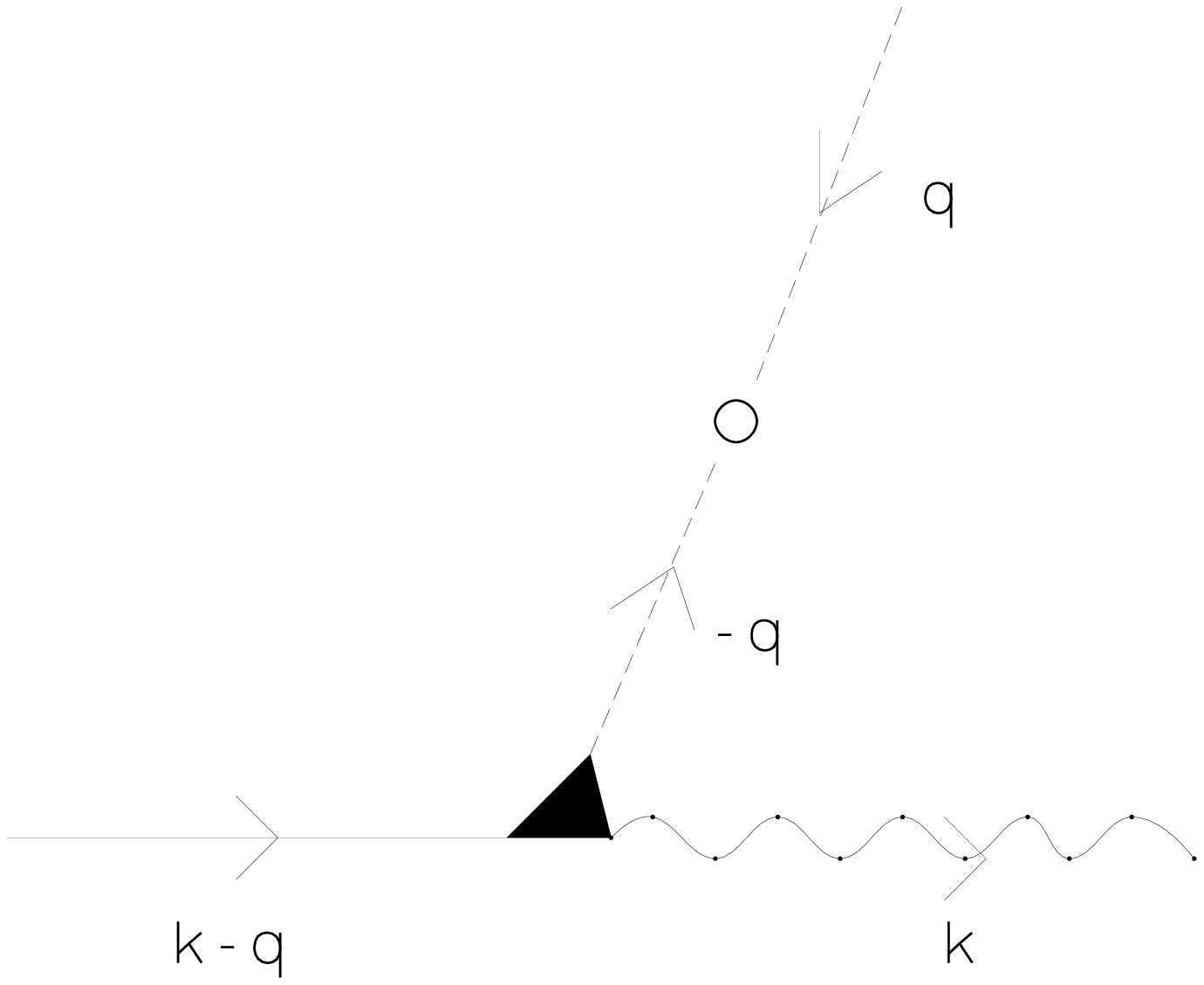}}
\caption{}
\end{figure}

\begin{figure}[htb]
\epsfysize=10cm
\centerline{\epsfbox{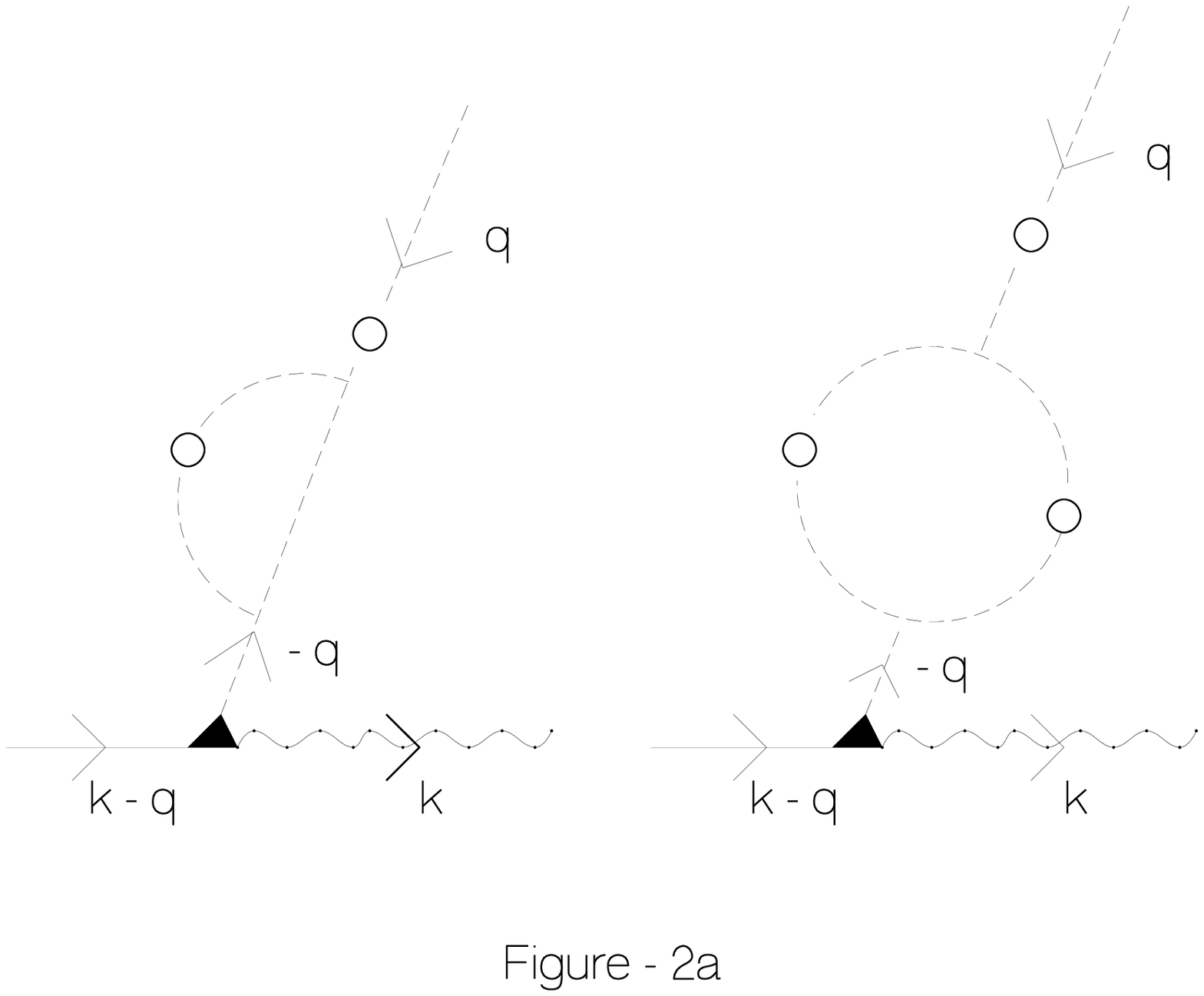}}
\end{figure}

\begin{figure}[htb]
\epsfysize=10cm
\centerline{\epsfbox{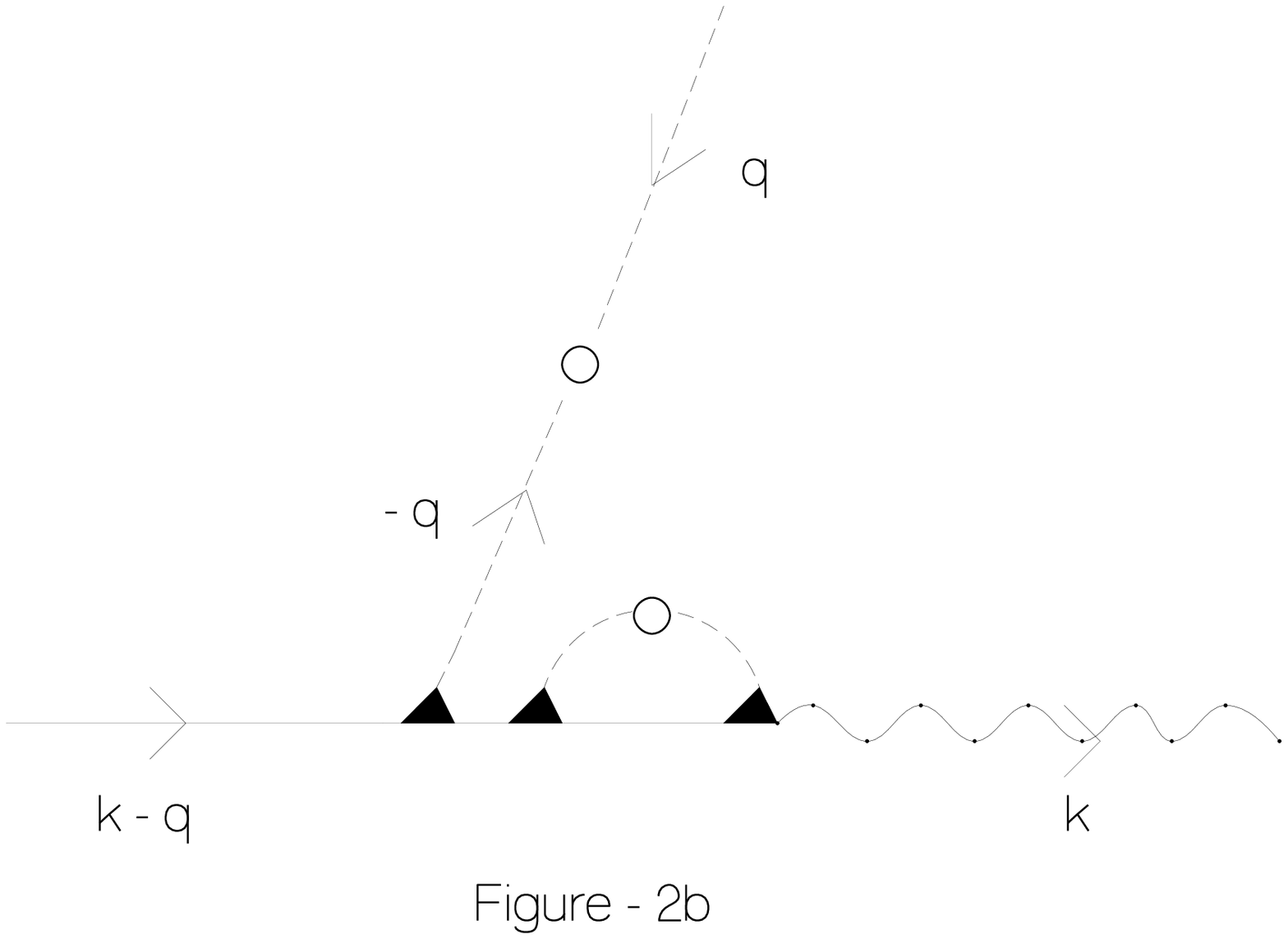}}
\end{figure}

\begin{figure}[htb]
\epsfysize=10cm
\centerline{\epsfbox{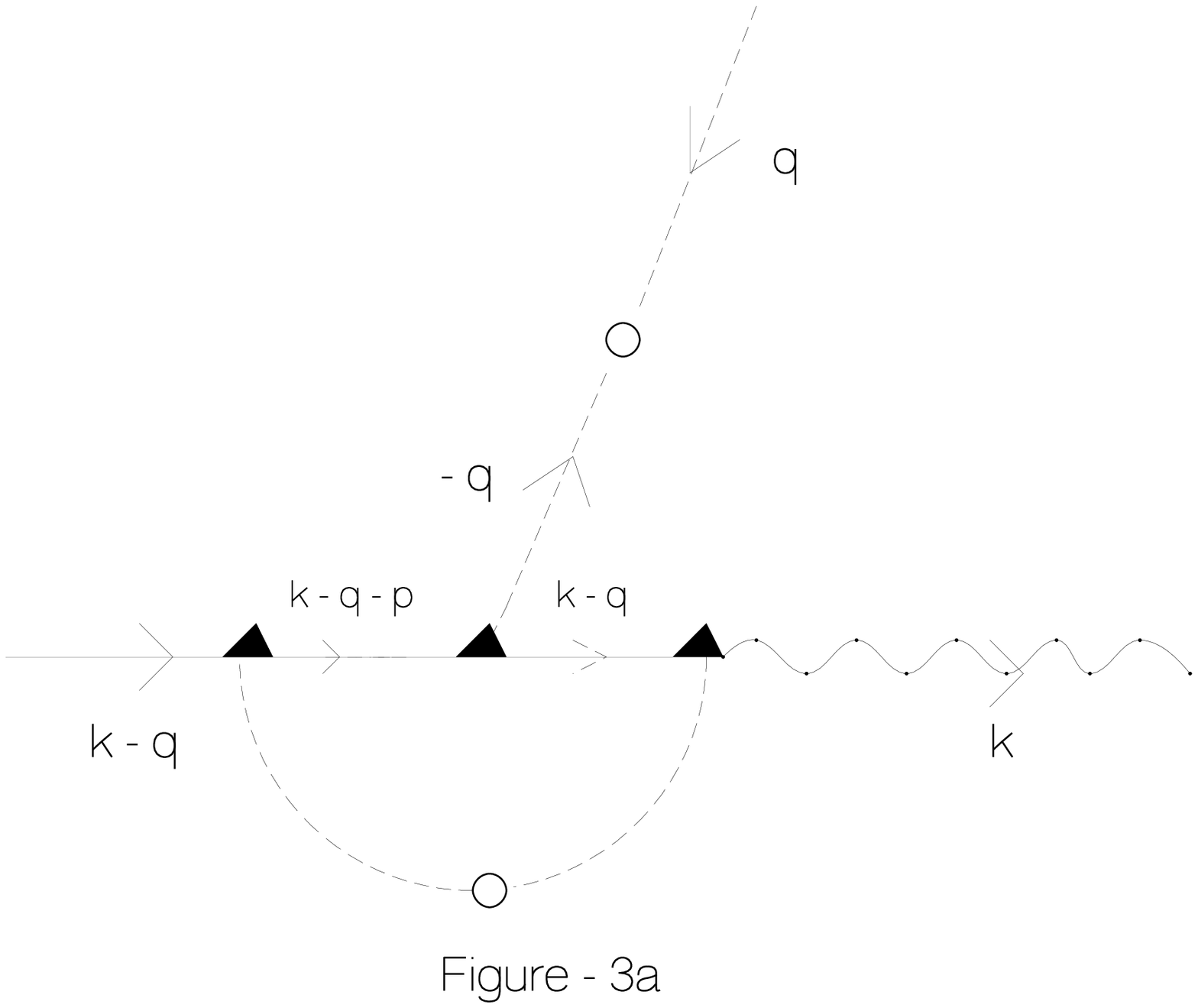}}
\end{figure}

\begin{figure}[htb]
\epsfysize=10cm
\centerline{\epsfbox{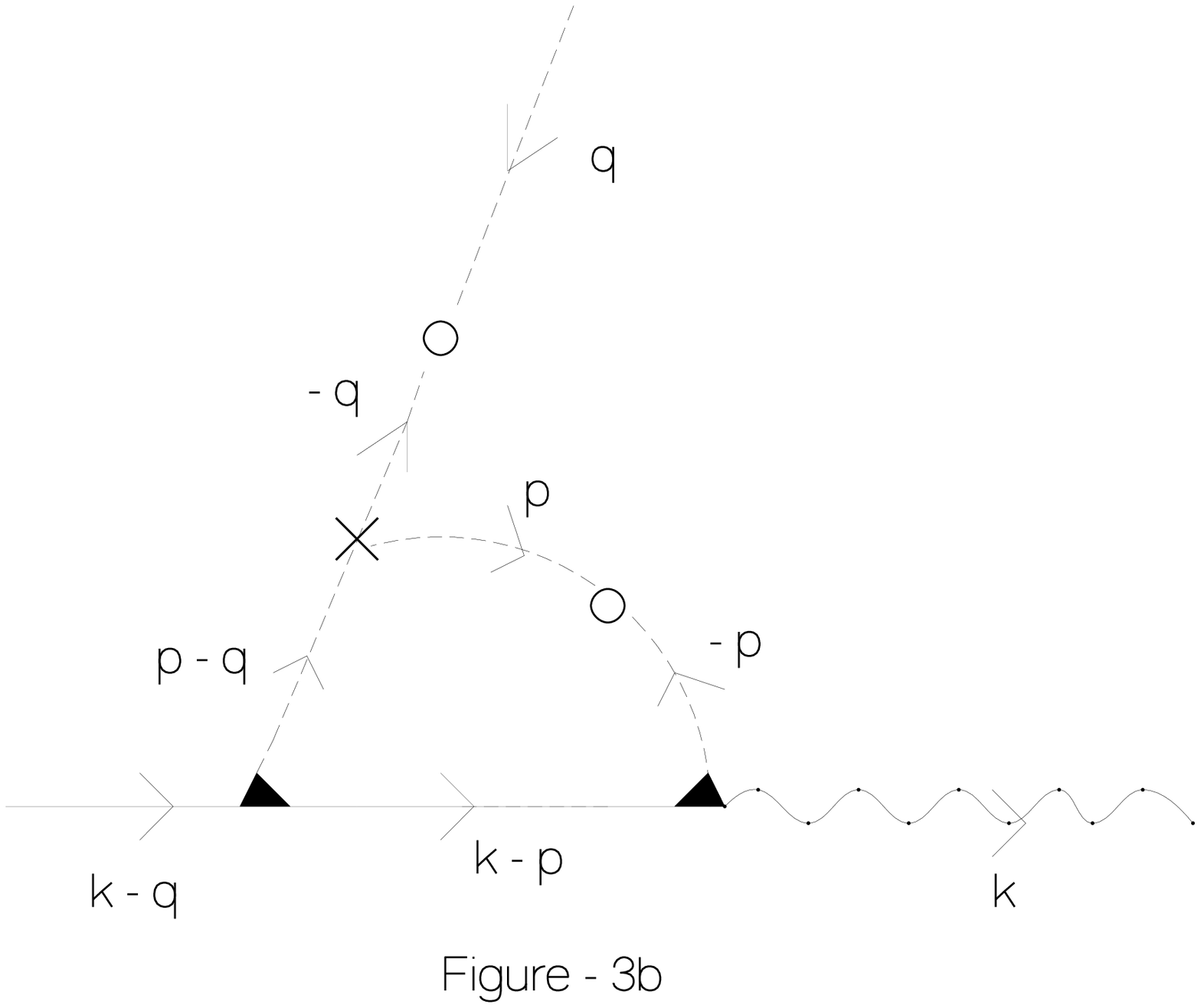}}
\end{figure}

\begin{figure}[htb]
\epsfysize=10cm
\centerline{\epsfbox{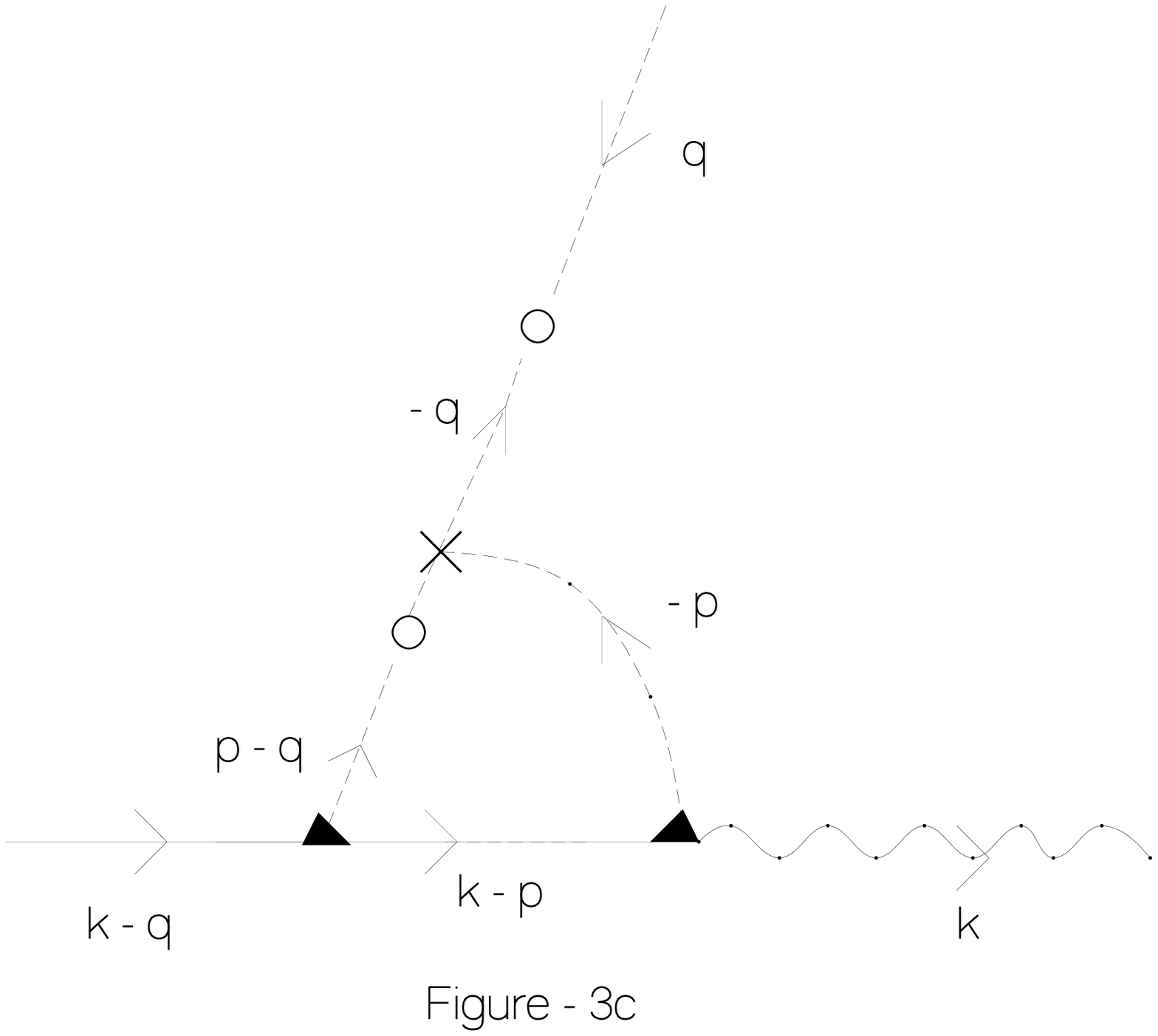}}
\end{figure}

\end{document}